\documentclass[12pt]{article}

\newcommand{\beg}{\begin{equation}}
\newcommand{\ene}{\end{equation}}

\begin{document}
\title{
\textsc{\bf Superluminal Dark Neutrinos}}

\author{Irina Ya. Aref'eva,
Igor V. Volovich
\\ $~~~$\\
\textsf{Steklov Mathematical Institute}\\
\textsf{Russian Academy of Sciences}\\
\textsf{Gubkin St. 8, 119991,  Moscow, Russia}\\
\emph{e-mails:arefeva@mi.ras.ru, volovich@mi.ras.ru}
}
\date {~}
\maketitle
\begin{abstract}
The OPERA collaboration has claimed the discovery of supeluminal
neutrino propagation. However the superluminal interpretation of the
OPERA result was refuted by Cohen and Glashow because it was shown
that such superluminal neutrinos would lose energy rapidly via the
bremsstrahlung of electron-positron pairs (arXiv:1109.6562).
We note that the superluminal interpretation is
still possible  if there exists a  new (dark) neutrino which can
propagate with a superluminal velocity and which couples with usual
neutrinos only via the mass mixing leading to
 neutrino oscillations.  It is supposed that the physical laws
 are invariant under rotations and translations in a preferred
reference frame. Two possible pictures to parameterize
 departures from Lorentz invariance  are discussed: a "conventional"
 tachyonic dark neutrino and the modification of the  Lagrangian
 by adding perturbations with the maximum attainable speed of the
 dark neutrino which is larger than the speed of light in vacuum.
We analyze also the MINOS and SN1987a data and show that they are
consistent with the conjecture that there exists the superluminal
dark neutrino.
\end{abstract}

\newpage
The special theory of relativity is a cornerstone of modern fundamental physics having the upper limit of velocities which is the velocity of light in vacuum. The OPERA collaboration has recently announced  the results about possible evidence for superluminal propagation of neutrinos \cite{OPERA}.
Specifically, the CNGS beam of muon neutrinos with mean energy of 17 GeV produced at CERN, travels about 730 km to the OPERA detector in the Gran Sasso Laboratory in Italy.
An early arrival time of the muon neutrinos with respect
to the one computed assuming the speed of light in vacuum of 60 ns is reported. This anomaly corresponds to a relative difference of the muon neutrino velocity with respect to the speed of light $\delta\equiv (v-c)/c=2.5\times 10^{-5}$.

Obviously such an astonishing claim requires of  course extraordinary standards of proof including confirmation by independent experiments.

The earlier MINOS experiment \cite{MINOS} reported a measurement of
$\delta=5\times 10^{-5}$ with lower neutrino energies peaking at 3 GeV. At the lower energy, in the 10 MeV range, a stringent limit of $|\delta|<2\times 10^{-9}$ was set by the observation of (anti) neutrinos emitted by the SN1987a supernova \cite{SN1987}.

There are various investigations of constraints on neutrino
velocities and  possible mechanisms for breaking the standard
Lorentz invariance motivated by the OPERA claim \cite
{Cacciapaglia:2011ax}--\cite{Sar}. It is known  that in the case of
superluminal propagation, certain otherwise forbidden processes are
kinematically permitted, even in vacuum, being analogues to
Cherenkov radiation. Cohen and Glashow \cite{CG} made an important
observation that especially the process of pair bremsstrahlung
\begin{equation}\label{pair}
\nu_{\mu}\longrightarrow\nu_{\mu}+e^+ +e^-\,.
\end{equation}
places a severe constraint upon the superluminal velocities.

From the neutral current weak interaction the authors
of \cite{CG} have computed the rate of pair emission
$\Gamma$ by an energetic superluminal neutrino
and $dE/dx,$ the rate at which it loses energy in
the high energy limit:
\begin{equation}\label{rates}
\Gamma=K^{'}G_F^2E^5\delta^3,\,\,\, dE/dx=-KG_{F}^2 E^6\delta^3.
\end{equation}
Here $K$ and $K^{'}$ are numerical constants and $\delta=v_{\nu}^2-1$
while the speed of light is set to unity.

Assuming $\delta$ is a constant one gets that neutrinos with
initial energy $E_0$, after traveling a distance $L$, will
have energy $E$ as given by the formula
\begin{equation}\label{energy}
E^{-5}-E_{0}^{-5}=5KG_{F}^2 L\delta^3\equiv E_{T}^{-5}\,.
\end{equation}
Therefore neutrinos with initial energy greater than the terminal
energy $E_T$ rapidly approach $E_T$ and the original beam would be
strongly depleted and spectrally distorted upon its arrival at the
Gran Sasso. Cohen and Glashow  conclude \cite{CG} that the
observation of neutrino with energies in excess of 12.5  GeV can not
be reconciled with the claimed superluminal neutrino velocity
measurement.

We suggest that the superluminal interpretation of the OPERA data is
 still possible if there is a  new (dark) neutrino which can propagate
 with a superluminal velocity and which couples with the usual neutrino
 only via the mass mixing leading to neutrino oscillations.

 We don`t assume the Yukawa coupling for the dark neutrino and the seesaw
 mechanism, so the dark neutrino is different
 from the sterile neutrino which could suffer from the difficulty
 with  pair bremsstrahlung.
 For the dark neutrino there is no
  process of pair bremsstrahlung (\ref{pair}) since
the dark neutrino sector couples with the standard model sector only
by means the mass mixing.

 To investigate possible violations
 of Lorentz symmetry  it is supposed that the physical laws are
 invariant under rotations and translations in a preferred
reference frame. Two possible pictures to parameterize
 departures from Lorentz invariance  are discussed: a "conventional"
 tachyonic dark neutrino and the modification of the  Lagrangian
 by adding perturbations with the maximum attainable speed of the
  dark neutrino which is larger than the speed of light in vacuum.
The last picture was considered by Coleman and Glashow for the
"usual" neutrino \cite{CG97,CG98}. They considered the case of
space-time translations along with exact rotational symmetry in the
rest frame of the cosmic background radiation, but allow small
departures from boost invariance in this frame. Perturbative
departures from Lorentz invariance are then  parametrized in terms
of a fixed time-like 4-vector.

 We analyze also the MINOS and SN1987a
data and show that they are consistent with the conjecture that
there exists the superluminal  dark neutrino.

Neutrino oscillations were predicted by  Pontecorvo \cite{Pont}.
It arises from a mixture between the mass and lepton flavor
(electron, muon or tau) eigenstates of neutrinos. A neutrino created
with a specific flavor can later be measured to have a different
flavor. The transformation from  the flavor to the mass eigenstates
is performed by means of the Pontecorvo-Maki-Nakagawa-Sakata (PMNS)
matrix.

Let us discuss first a toy model of particle oscillations when there
are only two scalar fields: the field $\phi$ with the usual mass $m$
and the superluminal (dark) tachyon field $\chi$. The Lagrangian is
\begin{equation}\label{lag}
{\cal L}=\frac{1}{2} (\partial_{\mu}\phi)^2-\frac{1}{2}m^2\phi^2+
\frac{1}{2}
(\partial_{\mu}\chi)^2+\frac{1}{2}M^2\chi^2+\lambda\phi\chi+{\cal
L}_{int}(\phi)
\end{equation}
where ${\cal L}_{int}(\phi)$ depends only on the field $\phi$. The
superluminal dark sector $\chi$ couples with the usual field $\phi$
only via the quadratic term $\lambda\phi\chi$. To quantize the
tachyon field $\chi$ we have to restrict ourself integration in the
momentum space over momenta greater than $M$, so we have violation
of Lorentz invariance.

There are particle oscillations described by this Lagrangian. If we
neglect the interaction term ${\cal L}_{int}(\phi)$ then one can go
from the flavor basis $(\phi,\chi)$ to the mass eigenstates basis
$(\varphi_1,\varphi_2)$ with the mass squares $(\kappa_{1}^2,-\kappa_{2}^2)$ by using the PMNS matrix:
\begin{equation}\label{bog}
\phi=\cos\theta \varphi_1+\sin\theta \varphi_2,  \,\,
\chi=-\sin\theta \varphi_1+\cos\theta \varphi_2.
\end{equation}
For the creation and annihilation operators it will be just the
Bogoliubov transformation.
The mean value of energy for the $\phi$-particles will be
\begin{equation}\label{entach}
E=\cos^2\theta \sqrt{p^2+\kappa_{1}^2}+
\sin^2\theta\sqrt{p^2-\kappa_{2}^2}
\end{equation}
and the mean velocity is
\begin{equation}\label{entachvvel}
v=\frac{p\cos^2\theta}{ \sqrt{p^2+\kappa_{1}^2}}+
\frac{p\sin^2\theta}{\sqrt{p^2-\kappa_{2}^2}}
\end{equation}
with a rather complicated dependence from energy $E$.

To describe  propagation of the $\phi$ particle we consider the
amplitude $<f|e^{-itH}|f>$ where the state $|f>$ is a tensor product of the
one particle state in the physical sector of the field $\phi$ and
(bare) vacuum in the dark sector of the field $\chi$.  The role of
the wave packets in the theory of particle oscillations is
considered in \cite{DMOS}. This amplitude describes particle
oscillations and the particle of the physical field $\phi$
propagates with the superluminal velocity due to the coupling with
the dark sector of the $\chi$-field. In this system there is
entanglement between two sectors and one could disentangle
$\phi$-particles by taking trace over the dark sector. Vacuum decay in the presence of tachyons was discussed in \cite{Zel}. Gravity with Lorentz violation and superluminal propagation in  various models with extra dimensions have been considered  in \cite{Rub}-\cite{JA}. Note also the string models with right neutrinos on the
branes \cite{IAEKTT}.

Instead of coupling with the tachyon field one could use the
Coleman-Glashow parametrization of the violation of Lorentz
invariance. In this case the field $\chi$ has a normal mass and one adds the term   $-\frac{\epsilon}{2}(\partial_i\chi)^2$ which violates the Lorentz invariance:
\begin{equation}\label{lagCG}
{\cal L}=\frac{1}{2} (\partial_{\mu}\phi)^2-\frac{1}{2}m^2\phi^2+
\frac{1}{2}
(\partial_{\mu}\chi)^2-\frac{\epsilon}{2}(\partial_i\chi)^2-
\frac{1}{2}M^2\chi^2+\lambda\phi\chi+{\cal
L}_{int}(\phi)
\end{equation}
It is tempting to speculate that the Lagrangians of this type could be obtained by the dimensional reduction in the Kaluza-Klein or superstring theories when instead of Minkowsky  we reduce to a spacetime with less symmetries. One can assume that  true fundamental constants live in ten dimensions and the four-dimensional fundamental constants,
including not only the Newton constant but also
the speed of light and even the Planck constant, can be obtained as approximated values depending on the compactification.

The minimal extension of the standard model Lagrangian ${\cal
L}_{SM}$ along of this discussion with a Majorana mass term looks as follows
\begin{equation}\label{smext}
{\cal L}={\cal L}_{SM} +\frac{m_{\alpha\beta}}{2}\bar{\nu}^{c}_{L\alpha}
\nu_{L\beta}+h.c.+{\cal L}_{Lor}(\chi)
\end{equation}
Here $\alpha,\beta =e,\mu,\tau,\chi$. The field $\chi$ now denotes the superluminal dark  Majorana spinor and the
quadratic term ${\cal L}_{Lor}(\chi)$ involves only the field $\chi$
and violates Lorentz invariance.

Similarly one can build couplings with the Dirac superluminal dark fields. Whether one can construct a gauge invariant
formulation of such theories it has to be seen.
$~~~~$
\\
\\
\newpage
\section*{ Acknowledgments}

The work is partially supported by grants  RFFI 11-01-00894, NS 8265.2010.1 (IA) and RFFI 11-01-00828-a  and NS 7675.2010.1 (I.V.).

\end{document}